\documentclass[11pt,twoside]{article}

\usepackage{asp2004}
\usepackage{epsf}
\usepackage{psfig}
\usepackage{lscape}

\begin{document}
\title{Recent observational progress in AM CVn binaries} %%% Fill in title
\author{G. Ramsay,$^1$ C. Brocksopp, $^1$ 
P. Groot, $^2$ P. Hakala, $^3$ H. Lehto, $^3$
T. Marsh, $^4$ R. Napiwotzki, $^5$ G. Nelemans, $^2$ S. Potter, $^6$
B. Slee, $^7$ D. Steeghs,$^8$ and K. Wu $^1$.}   %%% Fill in author names
\affil{$^1$Mullard Space Science Lab/UCL, Holmbury St.Mary, Dorking, UK\\
$^2$Dept Astrophysics, Radboud University of Nijmegen, The Netherlands\\
$^3$Tuorla Observatory, University of Turku, V\"ais\"al\"antie 20, Finland\\
$^4$Dept Physics, University of Warwick, Coventry, UK\\
$^5$Centre of Astrophysics Research, Univ of Hertfordshire, Hatfield, 
UK\\
$^6$South African Astronomical Observatory, Cape Town, South Africa\\
$^7$Australia Telescope National Facility, Sydney, Australia\\
$^8$Harvard Smithsonian Center for Astrophysics, Cambridge, USA\\
}    %%% Fill in author affiliations

\begin{abstract} 
We present the results of some recent research on AM CVn systems. We
present: X-ray/UV observations made using {\sl XMM-Newton}; the X-ray
grating spectrum of RX J1914+24; preliminary results of a search for
radio emission from AM CVn binaries, and discuss the strategy and
first results of the RATS project, whose main aim is to discover AM CVn
systems.
\end{abstract}

%%% MAIN BODY OF TEXT GOES HERE. CONSULT "INSTRUCTIONS FOR AUTHORS USING
%%% LATEX2E MARKUP", SECTIONS 2.3-2.6 FOR HELP WITH EQUATIONS, FIGURES,
%%% AND TABLES.

\section{AM CVn binaries}   

Most cataclysmic variables (CVs) consist of a white dwarf accreting
material from a late-type main sequence star via Roche lobe
overflow. However, in some CVs, the mass donor is hydrogen deficient -
either a semi-degenerate star or a white dwarf. These binaries are
called AM CVn systems or ultra-compact binaries since they have
orbital periods less than $\sim$70 mins.  Currently there are 17
confirmed systems with another 2 candidate systems. This paper
presents: the main results of a survey of these systems using {\sl
XMM-Newton}; an initial study of the high resolution X-ray spectrum of
RX J1914+24; the results of a search for radio emission from RX
J0806+15 and RX J1914+24, and we also present initial results of the
RApid Temporal Survey (RATS) project whose aim is to detect new AM CVn
systems.

\section{Observations of AM CVn binaries using {\sl XMM-Newton}}   

Until the launch of {\sl XMM-Newton}, AM CVn systems had been little
studied in X-rays - only 3 of the then known systems were detected in
the {\sl ROSAT} all-sky survey. With the greater sensitivity of the
EPIC X-ray detectors on board {\sl XMM-Newton,} coupled with the
optical/UV telescope (the OM) it was possible to study these systems
in way that was not previously possible.

One of our goals was to search for evidence for a coherent modulation
in the X-ray intensity of these systems. Of the 7 systems which we
have so far observed (excluding RX J0806+15 and RX J1914+24), none
show any evidence for pulsations in their light curve. This suggests
that we have not detected the spin period of the accreting white
dwarf, and implies that they do not have strong magnetic fields.  The
sources did, however, show variability (cf Figure 1).

A second goal was to determine the nature of their X-ray emission. We
found that they were best modelled using optically thin thermal plasma
models with highly non-solar abundances. On top of their hydrogen
deficiency, many systems showed significant nitrogen over abundances.

Our third goal was to determine their energy balance for the first
time. We showed that a large fraction of their accretion energy was
emitted at UV wavelengths. In particular, the UV luminosity was
greatest at shorter periods where the accretion disc dominates.
Moreover, with parallaxes for these systems now available (Roelofs et
al 2006) we showed that their observed luminosity agrees remarkably
well with the accretion luminosity predicted by gravitational wave
losses. Our results are presented in detail in Ramsay et al.\
(2005b, 2006c).

\begin{figure}
\begin{center}
\setlength{\unitlength}{1cm}
\begin{picture}(6,7.5)
\put(-3,-0.7){\includegraphics{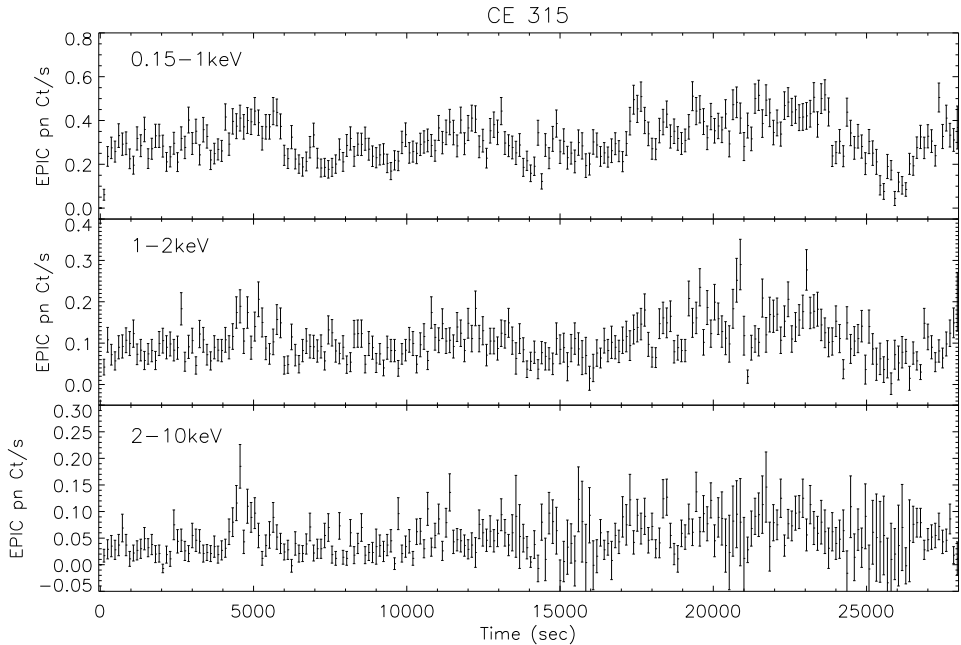}}
\end{picture}
\end{center}
\caption{The X-ray light curves of V396 Hya (CE 315) in 3 energy bands
obtained using {\sl XMM-Newton}.}
\label{rx0806-spin}
\end{figure}

\section{The candidate systems RX J0806+15 and RX J1914+24}

The X-ray sources RX J0806+15 (HM Cnc) and RX J1914+24 (V407 Vul) have
been the subject of much interest. If their periods (321 sec and 569
sec respectively) can be confirmed as their orbital period, then their
orbital periods will be the shortest among known binary systems. Here
we concentrate on two aspects - the nature of the X-ray spectrum of RX
J1914+24 and a search for radio emission from RX J0806+15 and RX
J1914+24.

The X-ray spectrum of RX J1914+24 is complex. It cannot be well fit
using a simple absorbed blackbody or absorbed optically thin thermal
plasma model. Rather, other model components (such as an absorption
edge or Gaussian line) have to be added to achieve good fits to the
{\sl XMM-Newton} EPIC spectra (Ramsay et al.\ 2005a, 2006a). It is not
clear if these extra components are physically realistic.

\begin{figure}
\begin{center}
\setlength{\unitlength}{1cm}
\begin{picture}(6,6.5)
\put(-2.6,-1.2){\includegraphics{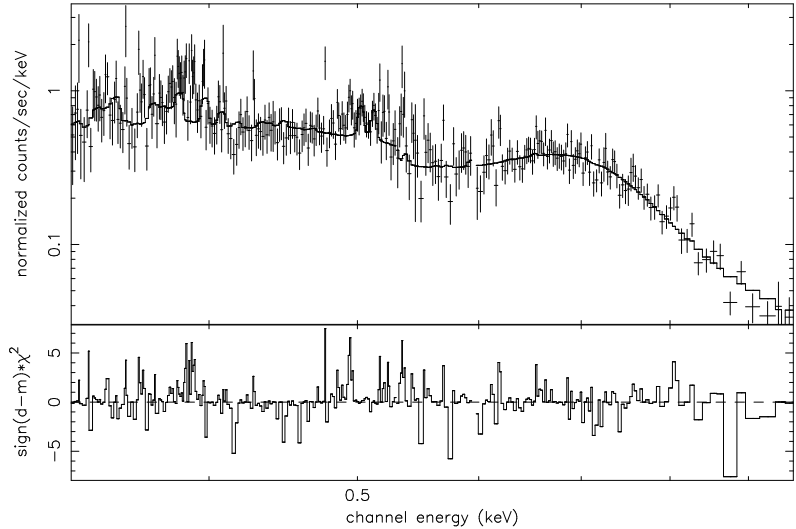}}
\end{picture}
\end{center}
\caption{The X-ray spectrum of RX J1914+24 taken using the {\sl
XMM-Newton} RGS detectors. The data from both detectors and data from
{\sl XMM-Newton} orbits 0718, 0721 an 0880 have been combined. The fit
was made using an absorbed thermal plasma model of variable abundances
plus a broad Gaussian feature centered at 0.68 keV.}
\label{rx1914-spec}
\end{figure}

To investigate this in more detail we have extracted a spectrum of RX
J1914+24 using the Reflection Grating Spectrometer detectors on-board
{\sl XMM-Newton}. We summed up the data from both RGS detectors and
from data taken in {\sl XMM-Newton} orbits 0718, 0721 and 0880. (RGS
data from orbit 0882 has various issues which we are still
investigating). We show the spectrum in Figure \ref{rx1914-spec}. We
find that an absorbed thermal plasma model plus broad Gaussian feature
centered at 0.68 keV gives a significantly better fit than either a
model which replaces the Gaussian component with an edge or a model
consisting of a blackbody (plus Gaussian or edge).

Determining if RX J1914+24 has an optically thick or thin X-ray
spectrum has important consequences for its luminosity. Ramsay et al.\
(2006a) show that the optically thin emission model gives a luminosity
2--3 orders of magnitude less than the optically thick based
models. If the emission layer is hotter than the optically thick
atmosphere beneath, then the spectrum would be expected to show
emission lines. There is some evidence for line features in Figure
\ref{rx1914-spec}. However, only a spectrum with a longer exposure
would be able to verify the presence of such lines. The physical
origin of the feature, to which we have fitted a Gaussian, is not
clear. Could it be some feature produced by a high magnetic field? For
instance broad absorption (rather than emission) features in X-ray
spectra of isolated neutron stars have been interpreted as due to
cyclotron features from which magnetic fields of 10$^{13}$ G have been
inferred (eg Haberl 2004).

The unipolar-inductor model (Wu et al.\ 2002) predicts that radio
emission should be produced due to electron-cyclotron maser emission
(Willes \& Wu 2004; Willes, Wu \& Kuncic 2004). If we observe the
system at an appropriate viewing which should detect strongly
circularly polarised radio emission. A search was made using the VLA
in Sept 2005 to detect emission from both RX J0806+15 and RX J1914+24
at 6cm. Each observation lasted a total of 3 hrs.  We did not detect
RX J1914+24 (3$\sigma$ upper limit of 42$\mu$ Jy/beam), but detected
RX J0806+15 at a level of 96$\mu$ Jy (a 5.0$\sigma$ detection). For
comparison we did not detect ES Cet (an accreting AM CVn system with
$P_{orb}$=10.4 min) using ATCA at 6cm (3$\sigma$ upper limit of
74$\mu$ Jy/beam) We have further observations of RX J0806+15 approved
with the VLA.

\section{The RATS project}

Less than 20 AM CVn systems are currently known. However, models
predict that many more systems should be easily detectable using
current instrumentation. For instance, Nelemans et al.\ (2004) predict
that nearly 4000 systems with periods shorter than 25 mins should be
brighter than $V$=22. These shorter period systems show modulation
periods in the optical band up to 0.1-0.2 mag.

There are a number of projects whose aim is to discover new AM CVn
systems.  The strategy of the RApid Temporal Survey (RATS) is to use
wide field cameras on medium sized telescopes to detect stellar
sources which vary in their optical intensity on periods shorter than
$\sim$80 min. A series of short ($\sim$30 sec) exposures are obtained
for a given field over the course of 2--3 hrs (Ramsay \& Hakala 2005).
Candidate systems are then subject to followup spectroscopy to
determine their nature.

\subsection{Results from our pilot study}

Our pilot study was obtained using the Isaac Newton Telescope on La
Palma with the Wide Field Camera. We obtained data of 12 fields which
gave a coverage of 3 square degrees. A total of 46 objects were found
to vary in their optical brightness. Details of our analysis and
results are given in Ramsay \& Hakala (2005). Many of the objects were
contact binaries (also known as W UMa stars). However, 4 sources
showed variability with periods on timescales less than $\sim$70
mins. Followup spectroscopy and photometry revealed their
nature. Three of these systems are SX Phe or dwarf Cepheid stars -
only $\sim$20 of these pulsating stars were previously known outside
stellar clusters. The object showing the shortest period, 374 sec, is
a rare pulsating sdB star (also known as EC 14027 stars). Its high
pulsation amplitude is consistent with its location at the cool end of
the EC 14026 instability strip (see Ramsay et al.\ 2006b for details).

\subsection{Latest results}

\begin{figure}
\begin{center}
\setlength{\unitlength}{1cm}
\begin{picture}(6,4.5)
\put(-4,-2.7){\includegraphics{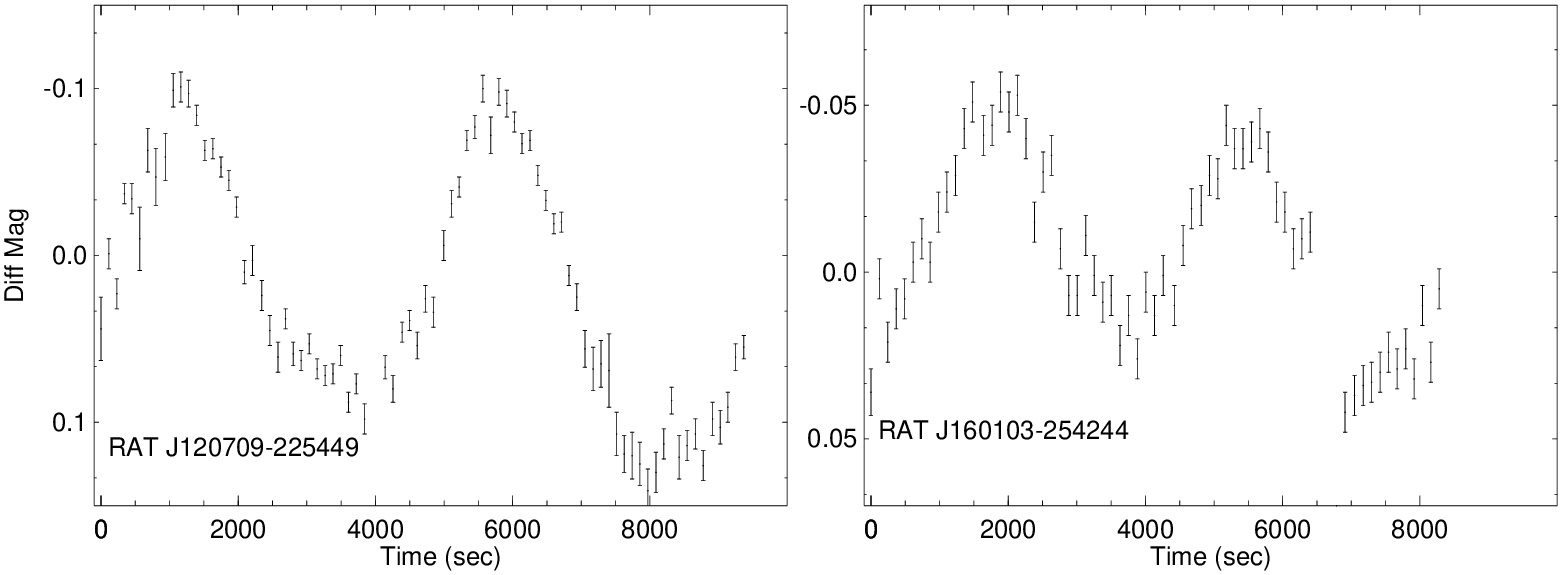}}
\end{picture}
\end{center}
\caption{The light curves of two systems discovered using the ESO 2.2m 
telescope. Followup spectroscopy is needed to determine their nature.}
\label{eso-rats}
\end{figure}

Since our pilot survey we have obtained data covering 12 fields using the 
ESO 2.2m telescope in Chile and 8 fields using the INT in La Palma, both 
in June 2005. Our analysis of the data from ESO is complete. Excluding those
fields which included globular clusters, we found over 150 new variable 
objects. Similar to our pilot survey, many of these are 
contact binaries (based on their light curves) and
around 10\% of these variables showed modulation periods
of less than $\sim$80 min (we show the light curves of two such systems in
Figure \ref{eso-rats}). A similar number of systems were found to be 
eclipsing systems. We have spectra of the brighter systems using the 
SAAO 1.9m telescope. 

It is also of interest to determine the number of ultra-compacts
located in globular clusters. This would give insight into how their
formation process was sensitive to the density of their
environment. We obtained data of M4 and M22 using the 2.2m telescope
in La Silla, and M71 using the INT.  We had to adapt our software to
suit the more crowded fields. However, the cluster centers (up to the
core radius) were too crowded for aperture photometry to perform well
-- a future study will develop tools which utilises profile fitting
photometry. We found a total of 24, 70 and 80 variable sources in
clusters M4, M22 and M71 respectively.  A significant proportion of
the sources are within the cluster tidal radius and therefore may be
cluster members. We have discovered more than 40 eclipsing systems and
more than 10 systems which show variations on periods less than 60
min.

\begin{figure}
\begin{center}
\setlength{\unitlength}{1cm}
\begin{picture}(6,7.5)
\put(-3,-0.5){\includegraphics{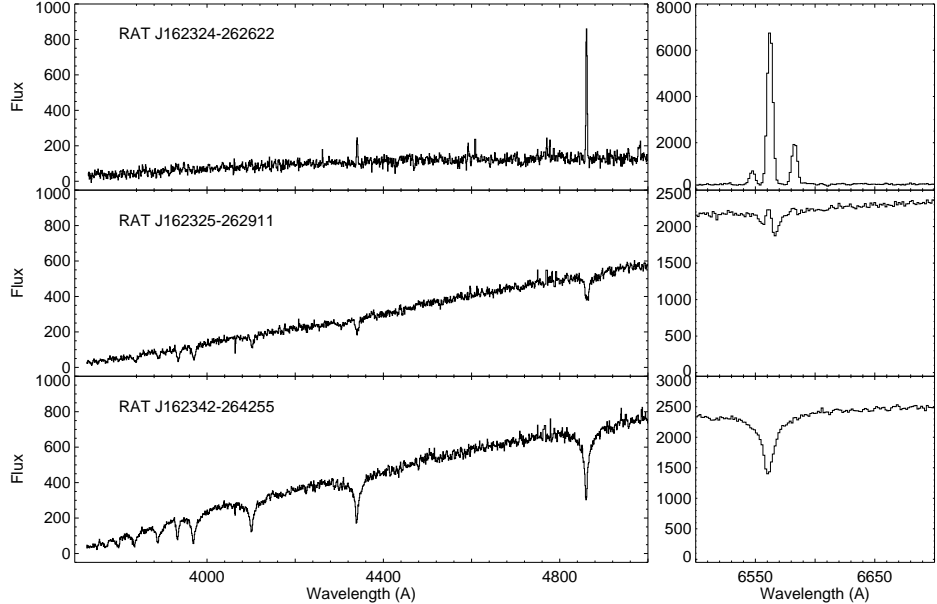}}
\end{picture}
\end{center}
\caption{The spectra of 3 variable sources discovered in the globular 
cluster M4 using data obtained from the 2.2m telescope at La Silla. 
The spectra were obtained using AAOmega and the AAT.}
\label{pn-spec}
\end{figure}

We have obtained spectra of the variable sources in M4 and M22 using
AAOmega at the Anglo Australian Telescope in the summer of 2006. In
the red arm we used the 385R grism and in the blue the 580V grism. The
seeing was moderate to poor so we did not obtain useful spectra for
the fainter targets.  There was no evidence for any ultra-compact
binaries in either cluster.  However, we did detect emission lines in
one system, RAT J162324--262622 which showed H$\alpha$-H$\gamma$ in
emission along with emission lines of forbidden N II on the blue and
red side of H$\alpha$ (Figure \ref{pn-spec}). Such a spectrum is more
similar to planatary nebulae.  RAT J162325-262911 shows emission in
the core of an absorption line at H$\alpha$, while RAT J162342-264255
shows deep Balmer absorption lines.

\subsection{Future strategy}

So far, we have not found any AM CVn systems in the RATS survey. At
face value this suggests that the models of Nelemans et al.\ (2001,
2004) overestimate the space density of AM CVn's. However, the size
of our survey is relatively small (6 square degrees have been
currently analysed in full) and secondly, the models predict that they
should be concentrated in the Galactic plane. Future field selections
will be strongly biased towards the Galactic plane
($|b|<10^{\circ}$). However, we have shown that our strategy is a good
means for discovering rare pulsating systems, some of which may
eventually become white dwarfs.

%%% THE BIBLIOGRAPHY
%%%
%%% CONSULT SECTION 3 OF "INSTRUCTIONS FOR AUTHORS" FOR HOW TO USE NATBIB.
%%% AUTHORS ARE ENCOURAGED TO USE EITHER THE "THEBIBLIOGRAPY" ENVIRONMENT
%%% BY UNCOMMENTING (DELETING THE "%" SYMBOL) THE COMMANDS BELOW, OR BY
%%% USING THE BIBTEX ENVIRONMENT. TO FIND OUT WHICH IS APPLICABLE TO YOUR
%%% CONTRIBUTION, CONSULT THE VOLUME EDITORS FOR YOUR PROCEEDINGS.
%%%

\end{document}